\documentclass[prb,twocolumn,showpacs]{revtex4}

\usepackage{graphicx}
\usepackage{dcolumn}
\usepackage{bm}
\usepackage{multirow}
\usepackage{amsmath,amssymb}

\begin{document}

\title{Quantum transport modeling of Fe/MgO/Fe magnetic tunnel junction with FeO$_{0.5}$ 
buffer layer: the effects of correlations \\}

\author{V. Timoshevskii}
\email{vladimir@physics.mcgill.ca}
\author{Yibin Hu}
\author{\'E. Marcotte}
\author{Hong Guo}

\affiliation{Department of Physics, McGill University, 3600 rue University, Montr\'eal, Qu\'ebec, Canada H3A 2T8}

\date{\today}

\begin{abstract}
We report \textit{ab initio} simulations of quantum transport properties of Fe/MgO/Fe trilayer structures with
FeO$_{0.5}$ buffer iron oxide layer, where on-site Coulomb interaction is explicitly taken into account by local
density approximation + Hubbard \textit{U} approach. We show that on-site Coulomb repulsion in the iron-oxygen
layer can cause a dramatic drop of the tunnel magnetoresistance of the system. We present an understanding of microscopic details of this phenomenon, connecting it to localization of the Fermi electrons of particular symmetry, which takes place in the buffer Fe-O layer, when on-site Coulomb repulsion is introduced. We further study the possible influence of the symmetry reduction in the buffer Fe-O layer on the transport properties of the Fe/MgO/Fe interface.
\end{abstract}


\maketitle


\section{Introduction}

Due to its high tunnel magnetoresistance ($TMR$), studying of Fe/MgO/Fe magnetic tunnel junctions (MTJ) has evolved into an active research area during last decade. Indeed, the TMR value of 180-250\%, reported by several research groups for MgO-based systems \cite{Yuasa2004,Parkin2004}, is the highest TMR, experimentally measured at room temperature. These high values of tunnel magnetoresistance make the MgO-based MTJs excellent candidates for spintronics applications: magnetic random access memory \cite{Moodera1995}, programmable logic elements \cite{Ney2003}, and magnetic read sensors.

\begin{figure}
\includegraphics[width=8cm]{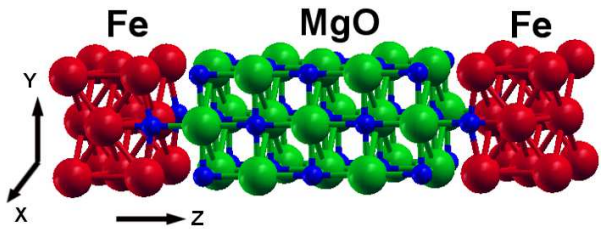}\\
  \caption{Atomic structure of the two-probe Fe/MgO/Fe device with Fe$_{0.5}$O buffer layer. The system is periodic in the directions ($X,Y$) parallel to the interface. The leads are semi-infinite in $Z$-direction.}
  \label{struct}
\end{figure}

Since the original theoretical prediction \cite{Butler2001} of large TMR effect in Fe/MgO/Fe interfaces, atomic-level simulations played a key role in understanding of unusually high TMR values in these systems, and a lot of progress has been achieved in this field. Butler \textit{et al} \cite{Butler2001}, using a Korringa-Kohn-Rostoker (KKR) technique, has provided an elegant explanation of physics behind the Fe/MgO/Fe magnetoresistance, connecting it with the match of the symmetry of the Bl\"och states in the electrodes with the symmetry of the evanescent states of the MgO barrier. Later, Waldron \textit{et al} \cite{Waldron2006b}, employing Non-Equilibrium Green's Function (NEGF) formalism, combined with Density Functional Theory (DFT), explained a voltage dependance of TMR in Fe/MgO/Fe interfaces. However, there is still an important issue which is the \textit{quantitative} understanding of physics behind Fe/MgO/Fe system: the absolute values of experimentally observed TMR ratios cannot be reproduced by \textit{ab initio} simulations if assuming ideal device models: these ideal models predicts much larger TMR than that has been observed. There may be a number of reasons behind the theory/experimental discrepancy such as oxygen vacancies in the MgO layer\cite{Mather2006,Youqi2010}, the existence of a FeO buffer monolayer\cite{Meyerheim2001,Butler2003}, other atomic defects\cite{Miao2008}, and interface roughness\cite{Waldron2006b}. These structural imperfections are clearly detrimental but may be overcome by improving the material quality and the fabrication process of the devices.
It is the purpose of this work to investigate a possible \emph{physical} reason - as opposed to the \emph{structural} reason, for the reduced TMR value as observed experimentally.

In particular, we theoretically investigate an important piece of physics that is related to the presence of a FeO buffer monolayer formed as a result of electrode oxygenation during the MgO deposition\cite{Meyerheim2001}. Calculations already demonstrated\cite{Butler2003} a reduction of TMR from that of the ideal device structure in the presence of the FeO layer, although even at 60\% oxygen concentration in the layer, the predicted TMR is still much greater than that observed experimentally. On the other hand, it is well known that \emph{bulk} FeO crystal is a Mott-Hubbard insulator whose electronic structure cannot be correctly predicted by Local Spin Density Approximation (LSDA) commonly used in density functional theory (DFT) calculations. This is due to an incorrect treatment of strong correlation in the FeO compound, as such LSDA predicts FeO to be metallic while in reality it is an insulator with a well-developed band gap. It is therefore very interesting to ask if a strong correlation exists in the single monolayer of FeO$_{1-x}$ that is at the interface of between Fe and MgO in the Fe/MgO/Fe trilayer so as to influence TMR. If it does, a localization of the Fe $3d$-states is expected that may lead to a substantial drop of transmission coefficients for both up- and down-spin channels and, as a result, to a significant change of TMR. To our knowledge, the only attempt to take into account correlation effects in FeO layer in Fe/MgO/Fe interface was taken by Mirhosseini \textit{et al} \cite{Mirhosseini2008}, who applied self-interaction correction (SIC) to LSDA to calculate spin conductance and magnetoresistance of Fe/FeO/MgO/FeO/Fe. These authors obtained a significant increase of the binding energies of Fe $3d$-states in the ``spin-up'' channel in FeO layer. As a result, it was found that SIC-corrected treatment moderately reduces the TMR ratio at energies below the Fermi level ($E_F$), from where the Fe $3d$ majority states are removed, while a 10\% increase of TMR at $E_F$ and above is observed. Since correlation effects may be important and it is a very complicated problem, further investigation along this line is warranted.

Here we present results of our first principles calculations of quantum transport properties in Fe/MgO/Fe interface with FeO$_{0.5}$ iron-oxide separation layer with the oxygen content close to the experimental one (60\%). We use a state-of-the-art quantum transport technique based on density functional theory (DFT) combined with Keldysh nonequilibrium Green's function (NEGF) formalism\cite{Taylor2001}. The correlation effects in FeO$_{0.5}$ layer are taken into account within the LDA+$U$ approach which explicitly adds the on-site Coulomb repulsion energy ($U$) for Fe $3d$-states directly to the DFT Hamiltonian. The value of $U$ is calculated from first principles for a particular atomic structure of the interface using a constrained LDA technique \cite{Anisimov91_1}. Our results demonstrate that the consideration of correlation effects in the FeO$_{1-x}$ layer leads to a dramatic drop of magnetoresistance in Fe/MgO/Fe interface, and brings it to the values comparable to that observed in experiments. The microscopic details of this substantial TMR reduction can be understood by analyzing the effects of electron localization on scattering states of different symmetry.

\section{Calculation details}
\subsection{Interface structure}
We consider a model Fe/MgO/Fe interface structure with five layers of magnesium oxide stacked between two semi-infinite Fe-bcc electrodes, oriented in the (100) direction (Fig. \ref{struct}). The system is periodic in the plane transverse to the interface direction. The unit cell vectors in this plane are rotated $45^\circ$ with respect to Fe-bcc unit cell, and their lengths equal $a\sqrt{2}$, where $a$ is the cell parameter of Fe-bcc lattice. This choice of the system geometry allowed us to model the interface FeO monolayer on each side of MgO slab with 50\% oxygen occupation (Fig. \ref{struct}),
which is close to 60\%, observed experimentally by Meyerheim \textit{et al}\cite{Meyerheim2001}.
The structure geometry was completely relaxed with respect to atomic positions and interlayer distances using highly accurate all-electron Linearized Augmented Plane Wave Method (LAPW) as implemented in Wien2k \cite{Wien2k} program package. We paid special attention to geometrical properties of the interface, as the earlier study of Waldron \textit{et al} demonstrated a substantial change in TMR, when the interface atoms are displaced by only 1\% of their bond length\cite{Waldron2006b}. Our theoretical optimized geometry showed a good agreement with results of X-ray diffraction experiments performed for the MTJs having 60\% oxygen content of the interface FeO layer \cite{Meyerheim2001}. In particular, the obtained Fe-O distances within the interface FeO layer and between FeO and MgO layers were found to be 2.03\AA~and 2.38\AA~respectively, which are in very good agreement with experimental data (2.03\AA~and 2.35\AA).

Having obtained the relaxed geometry of the interface, we proceed to calculate the electronic transport properties. We used the MatDCal program package \cite{Matdcal} which is a DFT-based code, employing a state-of-the-art quantum transport technique based on the Keldysh nonequilibrium Green's function (NEGF) formalism. The method uses standard norm-conserving
pseudopotentials \cite{TM-KB} and an \textit{s, p, d} double-zeta LCAO-type basis set. The NEGF-DFT formalism is excellently suited for self-consistent electronic transport calculations of multi-atom interfaces with non-periodic (open) boundary conditions. A detailed description of theoretical base of the method can be found in our previous publications
\cite{Taylor2001,Waldron2006a}. We found a $20\times20$ $K$-point mesh to be sufficient for self-consistent calculations of the two-probe device. Much denser ($10^3\times10^3$) sampling was further used for calculation of transmission coefficient $T$. A \textit{TMR} ratio was calculated at zero bias voltage using the following formula: $TMR=(T{_{\uparrow\uparrow}}-
T{_{\uparrow\downarrow}})/T{_{\uparrow\downarrow}}$, where $T{_{\uparrow\uparrow}}$ and $T{_{\uparrow\downarrow}}$ correspond to transmission coefficients calculated at the Fermi energy for parallel and anti-parallel spin configurations of iron leads.

\subsection{Correlation effects}

To study the effects of electron localization on electronic transport, we went beyond the standard NEGF-DFT formalism, and implemented the LDA+\textit{U} scheme into the MatDCal code. We followed the original works of Anisimov \textit{et al} \cite{Anisimov91_1,Anisimov91_2,
Anisimov93}, and details of our implementation of the LDA+\textit{U} method within NEGF formalism is summarized in the Appendix. According to the philosophy of LDA+\textit{U} approach, the atomic system is divided into the weakly-correlated part where the exchange-correlation is treated within LDA; and a strongly-correlated part where the Coulomb ($U$) and exchange ($J$) energies of interaction of two electrons on the same atom are explicitly taken into account. In the case of
oxidated Fe/MgO/Fe system, we treat the interface FeO$_{0.5}$ layer where we expect the electron localization to play a major role as a highly correlated subsystem and apply LDA +\textit{U} treatment to the Fe-$3d$ states of this layer.

To correctly apply the LDA +\textit{U} method, proper values of $U$ and $J$ parameters need to be supplied. These parameters are usually either taken from experimental data, or a series of calculations are performed for different values of $U$ and $J$ to check how the results depend on them. In our case, we have a single monolayer of FeO$_{0.5}$ sandwiched between an Fe electrode and the MgO barrier. To our knowledge, no experimental data about the on-site Coulomb repulsion energy
in this type of interfaces have been reported, hence we have calculated it from first principles. Note that the $U$ parameter, defined as the cost in Coulomb energy to place two electrons at the same site\cite{Anisimov91_1}, can be calculated by following the original ideas of Anisimov and Gunnarsson \cite{Anisimov91_1} within their particular implementation of the LAPW method, proposed by Madsen and Nov\'ak\cite{Madsen2005}. In our calculations, a supercell parallel to the interface plane was constructed, and the hybridization of $3d$-states of one of the Fe atoms in the FeO$_{0.5}$ layer with all other orbitals was switched off. By performing a series of self-consistent
calculations for this system with different values of occupation numbers of this artificially atomized $3d$-shell, we calculated the $U_{eff}=U-J$ parameter as a difference between the corresponding eigenvalues\cite{Madsen2005}. This way, we found that the effective Coulomb repulsion for Fe $3d$-electrons in the interface FeO$_{0.5}$ layer is about $4.6~eV$. This value is somewhat smaller than the one calculated by Anisimov \textit{et al} for bulk FeO crystal ($5.9~eV$) \cite{Anisimov91_2}. The reduction of $U$ in our system is expected due to the presence of the Fe electrode on one side of the FeO$_{0.5}$ layer: the free electrons in the Fe electrode increase the screening of the on-site Coulomb repulsion in the FeO$_{0.5}$.

\section{Results and Discussion}
\subsection{TMR ratio}

First, we performed full self-consistent calculations of the interface for Parallel (PC) and Anti-Parallel (APC) spin configurations of Fe leads for the cases of $U_{eff}=0$, as well as with ``turned on'' Coulomb repulsion. Further, for the case of each $U$-value, a contribution to transmission coefficient from spin-up and spin-down channels has been obtained, as well as a final value of $TMR$-ratio. The obtained results are summarized in Table \ref{table1}. The most striking result is the dramatic drop of the $TMR$ value for as soon as on-site Coulomb repulsion is taken into account for Fe $3d$-electrons in the FeO$_{0.5}$ interface layer.

\begin{table}[b]
\begin{center}
\caption{\label{table1}Calculated contributions to transmission coefficient $(\times 10^{-6})$ from different spin channels for the cases of $U=0$ and $U=4.6~eV$. The calculated $TMR$ ratio is also provided.}
\begin{ruledtabular}
\begin{tabular}{c c c c c }
Configuration & $T_{\uparrow}$ & $T_{\downarrow}$ & $TMR$ & $U_{eff}$ \\
\hline
PC & 95.57 & 0.81 & \multirow{2}{*}{1942\%} & \multirow{2}{*}{0.0~$eV$}\\
APC & 2.56 & 2.16 &       &         \\
\hline
PC & 2.08 & 0.47 & \multirow{2}{*}{60\%}   & \multirow{2}{*}{4.6~$eV$}\\
APC & 0.80 & 0.80 &       &         \\
\end{tabular}
\end{ruledtabular}
\end{center}
\vspace{-0.6cm}
\end{table}

For a pure LSDA calculation ($U_{eff}=0$), we obtain a $TMR$ of 1942\%, which is of the same order of magnitude as the results obtained in other similar calculations \cite{Waldron2006b}, as well as the LDA results obtained for oxigenated Fe/MgO/Fe interfaces ($\sim 1500\%$) \cite{Butler2003}. As soon as Coulomb repulsion of $4.6~eV$ is taken into account, the $TMR$ ratio drops dramatically down to $60\%$. To trace the origin of this drop, we analyze the transmission contributions to $TMR$ from different spin channels for $PC$ and $APC$  (Table \ref{table1}). We notice that the transmission in $APC$ for both spin channels reduces by a factor of $\sim 3$ when Coulomb repulsion is introduced. Even smaller is a reduction in spin-down channel for both PC and APC (factor of $\sim 1.7-2.7$). A really defining drop in transmission is observed in the spin-up channel for the case of PC. In this case the transmission coefficient drops by a factor of $46$, which constitutes a principal reason behind the dramatic reduction of $TMR$.

We should note that our LDA+$U$ results are somewhat opposite to those of Mirhosseini  \textit{et al}\cite{Mirhosseini2008}, who used the SIC-corrected LDA approach. We clearly see a significant drop of TMR ratio upon ``turning on'' correlations in FeO$_{0.5}$ layer, while Mirhosseini \textit{et al} did not observe significant changes of this quantity at energies equal or above the Fermi level. Further, our results demonstrate a principle reason of this drop - a dramatic reduction of transmission in the ``spin-up'' channel for PC, while the SIC correction mostly reduces the APC current.

\subsection{Scattering states}

To establish a more detailed physical picture, it is interesting to follow how the absolute square of the tunneling electron wavefunction (scattering state) changes along the tunneling direction. As the drop in $TMR$ ratio is caused by substantial reduction of transmission in spin-up channel for PC, we calculate the scattering state for this particular spin channel in PC. Only states at the $\Gamma$-point and at the Fermi energy are considered.

\begin{figure}[top]
\includegraphics[width=8cm]{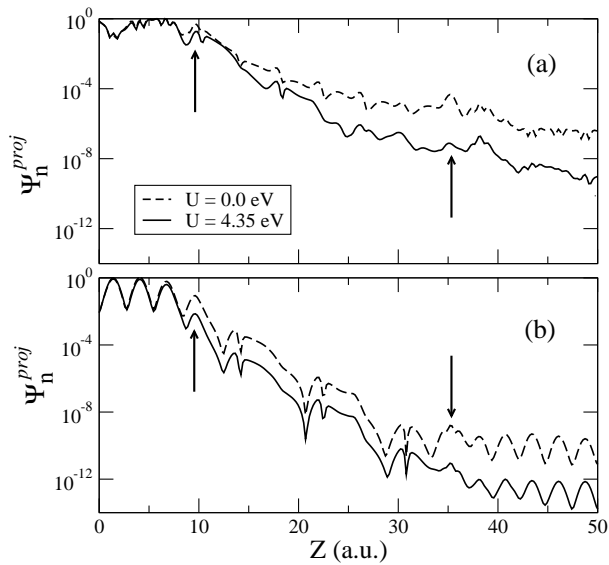}\\
  \caption{Distribution of squared electronic wavefunction (scattering state) across the
interface for two low-decay tunneling states of Fe/MgO/Fe interface with FeO$_{0.5}$ buffer layer. The wavefunction is integrated in $XY$ plane.
Calculations are done at the Fermi energy, for spin-up channel and at the $\Gamma$-point only. The arrows mark the positions of the FeO$_{0.5}$
layer on both sides of the interface.}
   \label{scatter}
\end{figure}

In their pioneering theoretical study, Butler \textit{et al} \cite{Butler2001} demonstrated that tunneling in the spin-up channel almost exclusively takes place through the state of $\Delta_1$ symmetry, while other bands, crossing the Fermi level of the Fe lead, rapidly decay in the MgO barrier. In our system we also observe two low-decay scattering states with similar profile, while
others decay faster in the MgO barrier. The presence of two of these states, as compared to Butler's setup where there is only one state of $\Delta_1$ symmetry, is explained by the choice of our unit cell, where there are two Fe atoms per layer in the direction perpendicular to tunneling. We plot the low-decay scattering states in Fig. \ref{scatter}. It can be noticed that these two states are substantially different. Even without the Coulomb repulsion $U$, there is a factor of $10^3$ difference in the
squared amplitude of these states after tunneling through the barrier.

To get insight into the physical origin of this difference, we calculate the spectral composition of the two scattering states, shown in figure \ref{scatter}, by projecting them on atomic orbitals of the interface Fe atoms. We were surprised to find that the orbitals, forming the $\Delta_1$-state ($s, d_{z^2}$) in the system without FeO layer (or with 100\%-populated FeO layer), were not the only ones forming these low-decay states in the case of our structure with FeO$_{0.5}$ iron oxide layer. Additionally, a significant contribution of $d_{x^2-y^2}$ Fe states was found in both scattering states, presented in Fig.\ref{scatter}. In fact, this contribution is comparable in value to the one of $d_{z^2}$ orbital, and even exceeds it for one of the scattering states. The calculations with $U=0$ showed that the contribution of $d_{z^2}$ and $d_{x^2+y^2}$ orbitals is respectively equal to 54\% and 30\% for scattering state (a), and 19\% and 56\% for scattering state (b), Fig.\ref{scatter}. We see that the scattering state (a) is mainly formed by the $d_{z^2}$ states, while the $d_{x^2+y^2}$ orbitals mostly contribute to the formation of scattering state (b). This difference in electronic structure is the underlying reason of the significantly different decay rate of these two scattering states. A much lower tunneling amplitude of state (b) can be understood as follows. The $d_{x^2+y^2}$ orbitals, which contribute mostly to this state, are oriented in the $XY$ plane (parallel to the interface), which reduces their overlap with $p_z$-orbitals of MgO layer. At the same time, the $d_{z^2}$ orbitals, which dominate in the formation of scattering state (a), are oriented along the $Z$ axis (perpendicular to the interface). This orientation increases their overlap with MgO oxygen states and therefore creates conditions for higher tunneling current.

\subsection{Electron localization}

\begin{figure}[b]
\includegraphics[width=7cm]{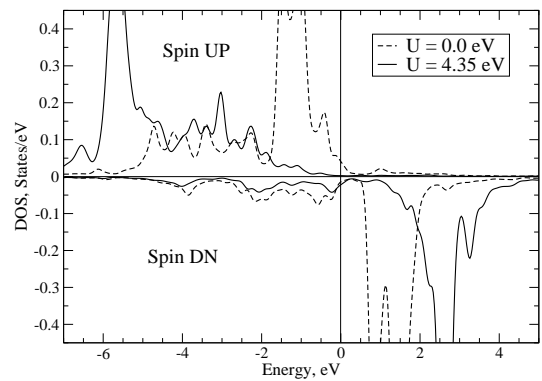}\\
  \caption{Density of electronic states for the Fe electrons of $d_{z^2}$ symmetry, calculated for the interface FeO$_{0.5}$ layer. Both results for calculations with $U=0.0~eV$ and with $U=4.35~eV$ are presented.}
   \label{dos}
\end{figure}

Let us now concentrate on the scattering state Fig.\ref{scatter}(a) which has the lowest decay rate, and therefore mostly contributes to the tunneling process. We see that ``turning on'' Coulomb repulsion $U$ gives rise to a factor of $\sim$300 drop in the squared amplitude of this state after tunneling through the barrier. To trace the physical origin of this substantial drop of transmission caused by the correlation effects in the FeO$_{0.5}$ interface layer, we study the density of electronic states (DOS) of the system. As more than a half or this scattering state (54\%) is contributed by Fe $d_{z^2}$ orbitals, we concentrate on the behavior of Fe $d_{z^2}$ states near the Fermi energy. Figure \ref{dos} shows the calculated DOS of $d_{z^2}$ states of Fe atoms in the FeO$_{0.5}$ interface layer. For the sake of comparison we plot the calculated DOS curves for $U=4.6~eV$ and $U=0.0~eV$ (standard LSDA). We see a substantial difference in behavior of Fe $d_{z^2}$-electrons for ``up'' and ``down'' channels upon ``turning on'' the Coulomb repulsion. For the spin-down states, the DOS at the Fermi level is practically unchanged, while for spin-up states it drops almost to zero when $U=4.6~eV$ is introduced. This is a well-known effect of LDA+\textit{U}: for the systems where correlation effects are strongly underestimated by standard LDA, the LDA+\textit{U} tends to split the DOS at the Fermi level into fully occupied and fully unnocupied peaks, significantly depleting DOS (or even forming the gap) at the Fermi level\cite{Anisimov91_2}.

\begin{figure}[]
\includegraphics[width=8cm]{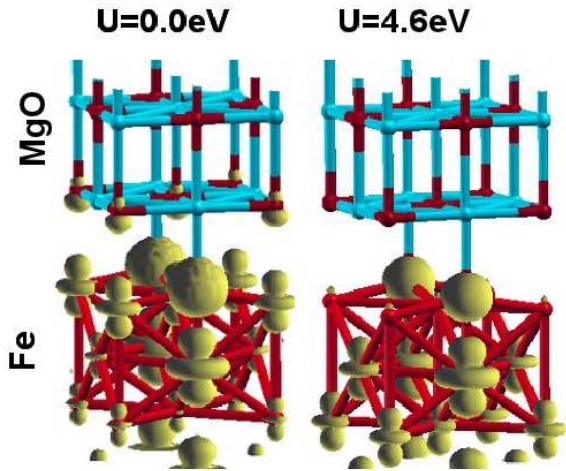}\\
  \caption{Real-space distribution of electronic density at the Fermi energy, calculated with $U=0.0~eV$ (left panel), and
with $U=4.35~eV$ (right panel) for oxidated Fe-MgO-Fe system. Only states, having $d_{z^2}$ symmetry, are taken into account for Fe layers when
calculating the density.}
\label{rho_dz2}
\end{figure}

The performed DOS analysis explains well the transmission data, summarized in Table \ref{table1}. Taking into account the correlation effects in the FeO$_{0.5}$ layer gives rise to a dramatic reduction of the electronic density of states at the Fermi level in the spin-up channel, which effectively ``turns off'' these electrons from the transmission process. In Table \ref{table1} this is reflected as a large drop in transmission in the spin-up channel when a non-zero \textit{U} is used in calculations. We decided to visualize this ``turn-off'' process for the $d_{z^2}$ Fe electrons in the FeO$_{0.5}$ layer by calculating their real-space distribution, and checking how this distribution is changed when the Coulomb repulsion is considered. In Figure \ref{rho_dz2} we plot the calculated real-space density of spin-up Fermi electrons for the interface part of the Fe/MgO/Fe system. We show only the electrons with $d_{z^2}$-symmetry of the wavefunction for all Fe atoms, while states with all symmetries are taken into account for O and Mg atoms. This symmetry-based plotting allows us to trace the behavior only of those electrons which participate in the tunneling process. The picture shows that in the case of standard LSDA calculations (left panel), the interface FeO$_{0.5}$ layer carries a significant density of $d_{z^2}$ Fermi electrons, and it is even comparable to the electronic density in the deeper layers of the Fe electrode. Moreover, these electrons couple with the oxygen $p$-electrons of the first MgO layer, creating a non-zero density of Fermi electrons on this layer. When the on-site Coulomb repulsion is turned on, the $d_{z^2}$ electrons with the Fermi energies are almost absent in the FeO$_{0.5}$ layer, as well as in the first layer of the MgO barrier. This picture illustrates that the proper consideration of the correlation effects in the interface FeO$_{0.5}$ layer leads to the localization of the Fe electrons with $d_{z^2}$ symmetry, which does not allow them to participate in the tunneling process and, as a result, is reflected as a dramatic reduction of the $TMR$ characteristics of Fe/MgO/Fe system.

\subsection{Low-decay states and the symmetry of FeO$_{0.5}$ layer}

Let us now go back to the question of spectral composition of the two low-decay scattering states, shown in Fig.\ref{scatter}. As we mentioned above, our calculations for the system with FeO$_{0.5}$ interface layer showed the presence of Fe $d_{x^2-y^2}$ orbitals in the low-decay scattering states: they contribute 30\% to the first lowest-decay state (Fig.\ref{scatter}a) and are mostly forming (56\%) the second one (Fig.\ref{scatter}b). The orbitals of this symmetry do not contribute to the $\Delta_1$-state in case of non-oxidated Fe/MgO/Fe interface\cite{Butler2001}, and in the following we shall try to understand the physical mechanisms of the influence of FeO$_{0.5}$ interface layer on the formation of the low-decay scattering states in Fe/MgO/Fe system.

Butler \textit{et al} \cite{Butler2001} brought on the idea of ``symmetry filtering'' which appeared to be the central point of the elegant physical explanation of $TMR$ effect in Fe/MgO/Fe trilayers. The authors demonstrated that Fe-bcc Bl\"och states of different symmetry show significantly different decay rates in the MgO barrier. The one with the smallest decay, which is able to couple to its
counterpart in the opposite Fe lead, is that of $\Delta_{1}$ symmetry. This state transforms as a linear combination of $s$, $p_z$, and $d_{z^2}$ orbitals. The analysis of Butler \textit{et al} is based on a perfect cubic symmetry of Fe-bcc, matching also perfect cubic unit cell of MgO (although rotated by $45^{\circ}$). In the case of 100\%-oxidated buffer layer (FeO) the situation does not change, as the point symmetry of all Fe sites is the same as in non-oxidated interface. However, the situation may change significantly if a low-symmetry FeO$_{1-x}$ layer is formed at the Fe/MgO/Fe interface. In this case the admixing of other electronic states to $\Delta_{1}$ band may take place (strictly speaking, this low-decay band can not be classified as $\Delta_{1}$ anymore.). This picture correlates well with our results for the interface with FeO$_{0.5}$ iron oxide layer. Although the distribution of oxygen in this layer is still ordered, the symmetry of Fe sites is changed. Therefore our structure can be a good model to test the effects of symmetry changes in FeO$_{1-x}$ layer on the composition of the low-decay scattering state.

To clarify the role of symmetry of FeO$_{0.5}$ iron oxide layer in formation of the low-decay states in Fe/MgO/Fe interface, we now take a different approach and perform a group theory analysis of the interaction of Fe lead states with the FeO$_{1-x}$ layer. In case of 100\%-populated FeO layer, all interface Fe atoms are occupying the positions with $C_{4v}$ point symmetry group. In this group the $d$-orbitals of iron and $p$-orbitals of oxygen of the first MgO layer transform according to the following irreducible representations: $A_1 \Rightarrow \{d_{z^2},p_z\}; ~E \Rightarrow
\{(d_{xz},d_{yz}), p_x, p_y\}; ~B_1 \Rightarrow \{d_{xy}\}; ~B_2 \Rightarrow \{d_{x^2-y^2}\}$. In this case the $A_1$ orbitals are forming the $\Delta_1$ band. This band is the only one having the admixture of O $p_z$-states, and has the lowest decay across the MgO barrier. The orbitals belonging to the two-dimensional representation $E$, form the $\Delta_5$ band. This band is doubly-degenerate and has a significantly higher decay in the MgO barrier. Finally, the orbitals belonging to $B_1$ and $B_2$ representations form $\Delta_{2'}$ and $\Delta_{2}$ bands, which are forbidden to interact with oxygen $p$-orbitals and show the highest decay in the MgO barrier. The results of this symmetry-based analysis are in excellent agreement with original physical picture presented by Butler \textit{et al} \cite{Butler2001}.

Let us now perform the same analysis for the case of our FeO$_{0.5}$ interface layer, where the point symmetry of Fe atoms is reduced. In this system the interface iron atoms have $C_{2v}$ point symmetry, and the interacting Fe $d$- and O $p$-orbitals belong to the following irreducible
representations \cite{symmetry}:  $A_1 \Rightarrow \{d_{z^2},p_z,d_{xy}\}; ~B_1 \Rightarrow \{(d_{xz}+d_{yz}), p_x\}; ~B_2 \Rightarrow
\{(d_{xz}-d_{yz}),p_y\};~A_2 \Rightarrow \{d_{x^2-y^2}\}$. We immediately see the striking difference: the $A_1$ representation now includes both $d_{z^2}$ and $d_{xy}$ orbitals while in the case of fully-symmetric FeO layer, the $d_{xy}$ orbitals were not allowed to interact with MgO
$p_z$-states by symmetry rules. It is important to note here that due to the $45^\circ$ rotation of the Fe electrode's unit cell in our calculation (as compared to the Fe-bcc unit cell), the calculated $d_{x^2-y^2}$ orbitals found to contribute significantly to the low-decay scattering states, are in fact the $d_{xy}$ orbitals in Fe-bcc unit cell \cite{symmetry}. We see that our results of theoretical symmetry-based analysis excellently agree with the results of our numerical simulations. If we now speak the language of the symmetry bands, coming out of Fe-bcc (100) surface, we observe that introducing of ordered FeO$_{0.5}$ interface layer in Fe/MgO/Fe interface leads to the mixing of $\Delta_1$ and $\Delta_{2'}$ bands and their interaction with $p_z$-states of oxygen in MgO layer. In this way, both $\Delta_1$ and $\Delta_{2'}$ bands contribute now to the formation of low-decay scattering state in MgO barrier, which change significantly the TMR properties of the system.

\begin{figure}[t!]
\includegraphics[width=8cm]{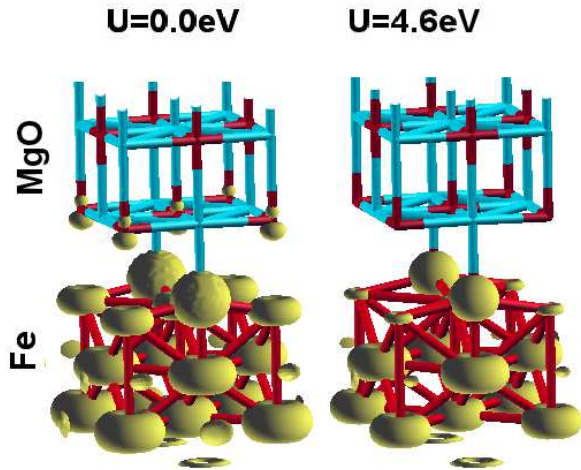}\\
  \caption{The same quantity as in Fig.\ref{rho_dz2}, but calculated with contribution from Fe $d_{x^2-y^2}$ states only.}
\label{rho_dx2y2}
\end{figure}

To complete our study, it is interesting to check if ``turning on'' the Coulomb repulsion influences interface Fe $d_{x^2-y^2}$ states in the same way as it collapses the $d_{z^2}$-states studied above. We calculated the real-space distribution of electrons with the Fermi energy for both cases of $U=0$ and $U=4.35eV$, where only the contribution from $d_{x^2-y^2}$ orbitals is taken into account. The results are presented in Fig.\ref{rho_dx2y2}. We see that a non-zero Coulomb repulsion gives rise to a significant collapse of the $d_{x^2-y^2}$ states in the FeO$_{0.5}$ layer. As a consequence, their coupling with oxygen states is dramatically reduced (notice the absence of charge in the first MgO layer for $U=4.35eV$ case, Fig.\ref{rho_dx2y2}, right panel), which leads to the $d_{x^2-y^2}$-states being practically ``switched off'' from the tunneling process. It is important to notice that the same effect should also take place in non-oxidated system, or in the system with 100\%-oxidated FeO layer, but in those cases the $d_{x^2-y^2}$ states are forbidden to interact with oxygen by symmetry and therefore this collapse should not influence in any way the tunneling characteristics of the interface.

\section{Conclusion}

In conclusion, we presented the results of \textit{ab initio} calculations of transport properties of Fe/MgO/Fe interface with FeO$_{0.5}$ buffer layer. By combining NEGF-DFT formalism with the LDA+\textit{U} approach, we explicitly took into account the on-site Coulomb repulsion in the Fe-O interface layer, where the \textit{U}-value was calculated from the first principles for a particular interface structure. Our results demonstrate a dramatic drop of TMR characteristic of the interface (down to $\sim$60\%), when Coulomb repulsion is taken into account in the FeO$_{0.5}$ buffer layer. This reduction is found to be mostly due to a dramatic drop of transmission in spin-up channel for parallel configuration of iron leads. By analyzing the distribution of scattering states across the interface, as well as a real-space distribution of the Fermi electrons with a particular symmetry of the wavefunction, we traced the physical origin of this transmission reduction. The results showed that for 50\%-oxidated buffer layer (FeO$_{0.5}$), this TMR drop is caused by a spacial collapse of Fe $d_{z^2}$ and $d_{x^2-y^2}$ electronic states at the Fermi level when non-zero $U$-value is taken into account. Finally, we used the methods of group theory and studied the general type of interaction between MgO layer and the buffer FeO$_{1-x}$ layer with different point group symmetry. The results show that the reduction of the symmetry of the buffer FeO$_{1-x}$ layer leads to the mixing of different symmetry bands of Fe-bcc bulk electrode, giving them the possibility to interact with MgO oxygen states and therefore to take part in the tunneling process.

\appendix
\section{LDA+\textit{U} approach in MatDCal code}

The main idea of the LDA+\textit{U} method is to replace the LDA $e-e$ Coulomb interaction energy of localized electrons in the energy functional of the system by a Hubbard-like term \cite{Anisimov91_1,Anisimov91_2,Anisimov93}:

\begin{equation}
\label{func}
 E_{LDA+U}[\rho_{\sigma}(\mathbf{r}),{n_\sigma}]=E_{LDA}[\rho_{\sigma}(\mathbf{r})]+E_U[{n_\sigma}]-E_{dc}[{n_\sigma}]
\end{equation}

Here $\rho_{\sigma}(\mathbf{r})$ is a real-space electronic density for spin state $\sigma$, $E_{LDA}[\rho_{\sigma}(\mathbf{r})]$ is a standard LDA functional, and $n_\sigma$ is a density matrix of the subsystem of correlated orbitals. The Hubbard-type term of this modified
functional can be written in the following way:

\begin{equation}
\label{Eu}
\begin{split}
E_U[{n_\sigma}]= & \frac{1}{2}U\sum_{m,m',\sigma}n_{m\sigma}n_{m'-\sigma}+  \\
                 & + \frac{1}{2}(U-J)\sum_{m\neq m',m',\sigma}n_{m\sigma}n_{m'\sigma}
\end{split}
\end{equation}

Here we assume that the on-site Coulomb ($U$) and exchange ($J$) interaction energies are independent of quantum numbers $m$. This expression also takes into account exchange: the interaction energy is $U-J$ for electrons with the same spin $\sigma$, while it is $U$ for electrons
with different spin. The last term in (\ref{func}) is a so-called ``double counting'' correction whose role is to subtract from the functional the averaged electron-electron interaction in the correlated subsystem, which has already been taken into account within LDA:

\begin{equation}
\label{dc}
E_{dc}[{n_\sigma}]=\frac{U}{2}N(N-1)-\frac{J}{2}\sum_{\sigma}N_{\sigma}(N_{\sigma}-1)
\end{equation}
where $N$ is the total number of electrons in correlated subsystem, and $N_{\sigma}$ is the total number of electrons with particular spin $\sigma$:
\begin{equation}
N_{\sigma}=\sum_{m}n_{m,\sigma};\,\,\,\,N=N_{\sigma}+N_{-\sigma}
\end{equation}

Now, by inserting (\ref{Eu}) and (\ref{dc}) in (\ref{func}), and employing the following expression for the total number of particles:
\begin{equation}
N^{2}=\sum_{m,\sigma}n_{m\sigma}^2+\sum_{m\neq m',m',\sigma}n_{m\sigma}n_{m'\sigma}+\sum_{m,m',\sigma}n_{m\sigma}n_{m'-\sigma}
\end{equation}
we arrive to a simple and computationally efficient expression for LDA+\textit{U} functional:
\begin{equation}
 E_{LDA+U}=E_{LDA}+\frac{1}{2}(U-J)(N-\sum_{m,\sigma}n_{m\sigma}^2)
\end{equation}

Taking the functional derivative leads to the following expression for the orbital-dependent potential:
\begin{equation}
\label{pot1}
 V^{LDA+U}_{m\sigma}=V^{LDA}+\frac{1}{2}(U-J)(1-2n_{m\sigma})
\end{equation}

As we notice, this form of LDA+\textit{U} potential depends on a single parameter which we call the ``effective'' Coulomb repulsion $U_{eff}=U-J$. This quantity can be calculated from the first principles for a given atomic system using a constrained LDA approach of Anisimov and Gunnarsson\cite{Anisimov91_1,Madsen2005}.

To employ this scheme in localized basis set which the MatDCal package uses, we write the orbital-dependent part of the potential in operator form:
\begin{equation}
 \hat{V}_{m\sigma}=V_{m\sigma} |m\sigma \rangle \langle m\sigma |
\end{equation}
where $V_{m\sigma}$ is the second term in (\ref{pot1}), and $|m\sigma \rangle \langle m\sigma |$ is a projection operator on the atomic orbital basis function corresponding to the localized state, for which LDA+$U$ treatment is applied. In this form we add the LDA+$U$ potential in a
straightforward manner to the MatDCAL software where the Hamiltonian matrix elements are calculated. Having computed the matrix elements, we follow the standard computational scheme of NEGF method in localized basis set representation. The technical details of this implementation
can be found in several previous publications \cite{Taylor2001, Matdcal,Marcotte}.

\bibliography{femgo_ldau}

\begin{thebibliography}{23}
\bibitem[{sym()}]{symmetry}
\bibinfo{note}{Due to the semiinfinite nature of the iron lead, we assume the
  geometry of the electronic wavefunction, entering the FeO$_{0.5}$ layer, to
  be the same as in the lead (Fe-bcc cubic lattice). However, the square
  lattice of FeO$_{0.5}$ layer, as well as the MgO lattice, are $45^\circ$
  rotated with respect to Fe-bcc unit cell. This fact is taken into
  consideration when we classify Fe states, coming out of the lead, in the new
  symmetry of FeO$_{0.5}$ and MgO layers. The most important in this context is
  that the $d_{x^2-y^2}$ orbital, which significant contribution to the
  scattering states was obtained by numericall computations, is in fact the
  $d_{xy}$ orbital in Fe-bcc lattice.}
\end{thebibliography}

\end{document}